# Coils and the Electromagnet Used in the Joule Balance at the NIM

Zhonghua Zhang, Zhengkun Li, Bing Han, Yunfeng Lu, Shisong Li, Jinxin Xu and Gang Wang

*Abstract*—In the joule balance developed at National Institute of Metrology (NIM), the dynamic phase of a watt balance is replaced by the mutual inductance measurement in an attempt to provide an alternative method for the kg redefinition. But for this method a rather large current in the exciting coil, is needed to offer the necessary magnetic field in the force weighing phase, and the coil heating becomes an important uncertainty source. To reduce coil heating, a new coil system, in which a ferromagnetic material is used to increase the magnetic field was designed recently. But adopting the ferromagnetic material brings the difficulty from the nonlinear characteristic of material. This problem can be removed by measuring the magnetic flux linkage difference of the suspended coil at two vertical positions directly to replace the mutual inductance parameter. Some systematic effects of this magnet are discussed.

## I. INTRODUCTION

THE joule balance method was proposed by NIM in 2006 to measure the Plank constant and redefine the kilogram [1]. The key point of this method is to avoid the dynamic measurement during the coil movement for the watt balance method [2], [3], to try giving another method for the kg redefinition because comparison of results from different methods could get more information. The joule balance requires a rather large current in the exciting coil to provide the necessary magnetic field for the weighing phase. Thus the coil heating becomes an important uncertainty source. For example, in the initial prototype of joule balance, a model coil system was constructed as shown in Fig. 1 (a). It is found that a 250 mA current should to be used for both the exciting and the suspended coil to generate a 2 N electromagnetic force. The power consumed in coils in this case was more than 120 W and leading to a 60 ppm uncertainty due to deformation of the coil caused by heating and forces produced by rising warm air. Later a more compact coil as shown in Fig. 1 (b) was used [4]. The power in coils apparently decreased to 40 W, and the magnetic force increased to 3 N. The uncertainty due to the coil heating was decreased to about 9 ppm.

It was also considered to use a superconducting coil to decrease the coil heating. But experiments showed that the Meissner effect of superconductor can bring a systematic uncertainty. Moreover, a superconducting system is expensive and difficult to run. Finally, the superconducting system has been given up.
In this paper, a new magnet with coils and the ferromagnetic material core is presented to serve the new generation joule balance. The new design reduces the coil heating to 7.5 W and increases the magnetic field by a factor of more than 10, therefore, the measurement uncertainty for the Planck constant is expected to be reduced further. A known problem for the new designed coil magnet is the nonlinearity due to the ferromagnetic material. To solve this problem, we propose to replace the mutual inductance measurement with a flux linkage measurement. Theoretical analysis and optimization of the secondary current effects in the weighing phase are discussed.

Manuscript received Aug. 26, 2014. This work was supported in part by the National Natural Science Foundation of China (51077120), and the National Department Public Benefit Research Foundation of China (201010010).

Zhonghua Zhang is with National Institute of Metrology, Beijing100029, China. Phone: +86-10-64524211, e-mail: zzh@nim.ac.cn.
Zhengkun Li, Bing Han and Yunfeng Lu are with National Institute of Metrology, Beijing 100029, China. Shisong Li and JinxinXu are with the Department of Electrical Engineering, Tsinghua University, Beijing 10084, China. Gang Wang is with the School of Instrument science and Opto-electronic Engineering, Beihang University, Beijing 100083, China.

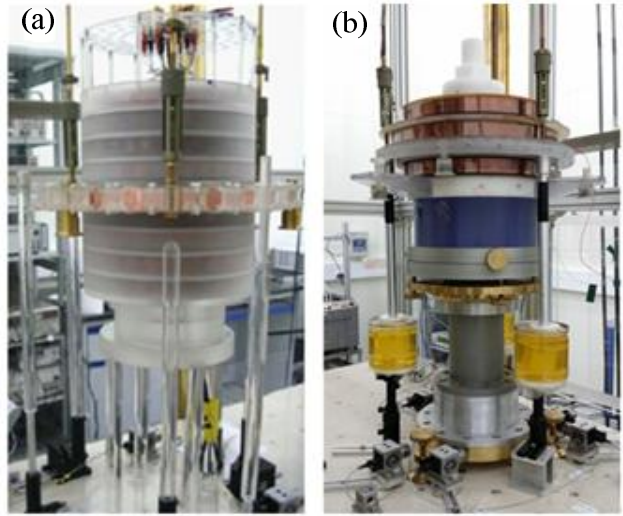

Fig. 1. Historical coils used in joule balance at NIM. (a) was the first coil set and (b) was the second coil set. In the first coil set, the fixed coil contained 8 coils in series (upper 4 and lower 4 were oppositely connected) to obtain a flat magnetic profile along the suspended coil measurement interval. In the second coil set, only two fixed coils were used to produce a flat profile, and the suspended coil was operated between them to generate a stronger magnetic force.

## II. Design of the Electromagnet

It is shown in the historical measurement data that turning from the first model coil set in Fig. 1 (a) to a compact coil set in Fig. 1 (b), the uncertainty due to the coil heating is reduced considerably. The reason is that in Fig. 1 (b) the distance between the exciting coil and the suspended coil is much smaller than that in Fig. 1 (a). The efficiency of the current in the exciting coil of Fig. 1 (b) producing useful magnetic field is better than that of the first coil set in Fig. 1 (a). According to this idea, it may be considered that, if the ferromagnetic material is used to short the magnetic path between the exciting coil and the suspended coil, the circumstance may become better and the coil heating problem could be improved further.

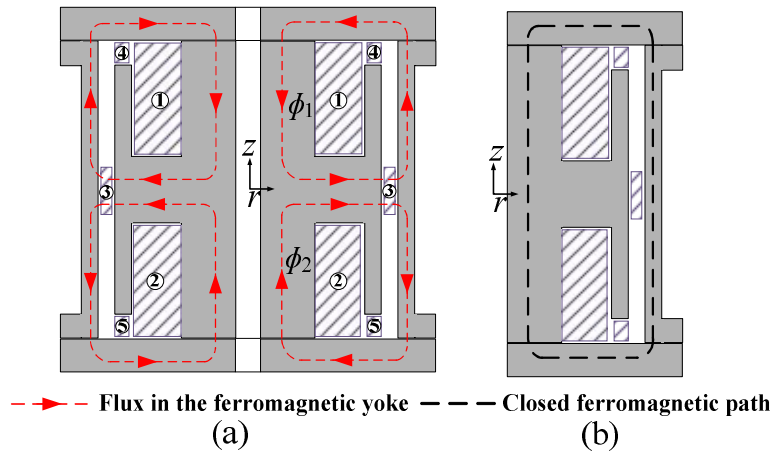

Fig. 2. Sectional view of the magnet with coils and the core made of ferromagnetic material. The magnet has a symmetrical structure and the numbers denote: 1—upper exciting coil, 2—lower exciting coil, 3—suspended coil, 4—upper compensation coil, and 5—lower compensation coil. The ferromagnetic yokes are made from DT4C. (a) Red routes denote the main magnetic flux. (b) Black dotted line denotes the closed ferromagnetic path.

Fig. 2 (a) shows the designed electromagnet with a symmetrical structure in both the vertical and the radial directions. An obvious feature of the magnetic circuit in Fig. 2 (b) is that there is a closed ferromagnetic path of high permeability formed from the inner vertical core, outside cylinder, upper and lower cover. A small magnetic motive force will cause a considerable additional magnetic flux in this closed magnetic path of high permeability. Therefore, two fixed exciting coils 1 and 2 must be with the same ampere-turns but connected in opposite direction, then the net magnetic motive force in the closed ferromagnetic path in Fig. 2 (b) is zero and the undesired additional flux is also zero. At the same time, two flux loops, noted as $\Phi_1$ and $\Phi_2$ respectively produced by exciting coils with inversed directions are formed at the upper and lower parts of the gap. As a result, a horizontal magnetic field appears in the gap. Note that the flux value is mainly depended on the reluctance of the air gap.

In Fig. 2 (a), number 3 denotes the suspended coil in the air gap. A problem is, when the suspended coil is carrying current, it would also produce a magnetic motive force for the closed ferromagnetic path of high permeability shown in Fig. 2 (b) and a considerable additional flux will appear in the magnetic path. Therefore, coils 4 and 5 are arranged for compensating this magnetic motive force of the suspended coil, named the upper compensation coil and the lower compensation coil respectively. The number of turns of every compensation coil equals the half of the turns of the suspended coil 3. Two compensation coils are connected in series in the same direction, and oppositely connected with the suspended coil 3 as the whole secondary coil.

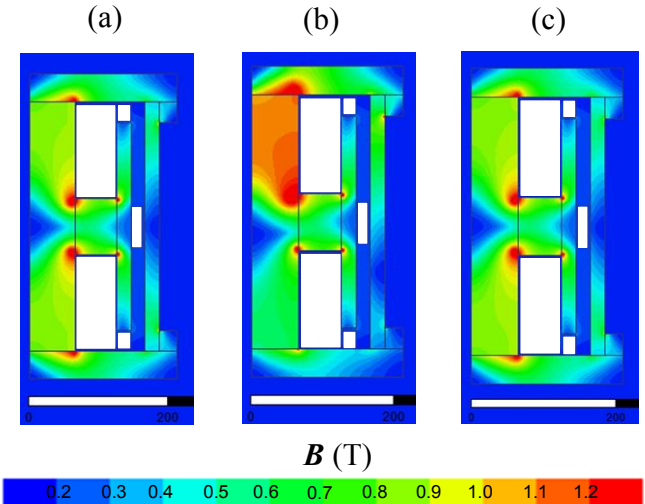

Fig. 3. Demonstration of the function for compensation coils. (a) The magnetic field generated by currents in exciting coils only. (b) The magnetic field given by currents in exciting and suspended coils, but current in compensation coils is 0. (c) The magnetic field given by currents in exciting and suspended coils, also compensation coils. Note that in three cases of the simulation a 0.05mm installation gap between the cover and the main yoke (inner vertical core/outside cylinder) is applied.

In Fig. 3, the function of the compensation coils is demonstrated. Fig. 3 (a) is the magnetic field offered by the current of exciting coils only and Fig. 3 (b) is the case when current of the suspended coil is applied without any current in compensation coils. It is seen that the magnetic field is distorted notably due to the additional magnetic field from the suspended coil and an additional error would appear. But in reality, a same current is applied in both the suspended and compensation coils simultaneously, and the magnetic field distribution as shown in Fig. 3 (c) is similar to that in Fig. 3 (a). It shows the current in compensation coils will remove the additional error. Therefore, the function of compensation coils is essential. Of course, these compensation coils will bring some additional current heating. But it is only about 1/40 of the current heating from the exciting coils and is not very important.

The ferromagnetic material used to construct the magnet is a kind of pure electric iron (type DT4C). Note that the DT4C is different from the yoke material applied in the watt balance, e.g., the low carbon steel. The measured $\mu \sim H$ curve (after annealing) is shown in Fig. 4. Compared with the low carbon steel, the DT4C material is employed from following two considerations.

(1) Its maximum magnetic permeability is about 8300, which is much higher than the steel used in a watt balance, e.g., NIST-4, the 4th generation of watt balance in the National Institute of Standard and Technology (NIST) [5]. This would lead to a better boundary condition of the air and the yoke. As a result, the vertical component of the magnetic flux density in the air gap is reduced, which benefits an easier alignment procedure.

(2) The magnetic field at the maximum magnetic permeability is 120 A/m, which is much smaller than that of the low carbon steel (typical 400 A/m - 500 A/m). It means that a less current in the exciting coil is required to reach the maximum $\mu_r$.

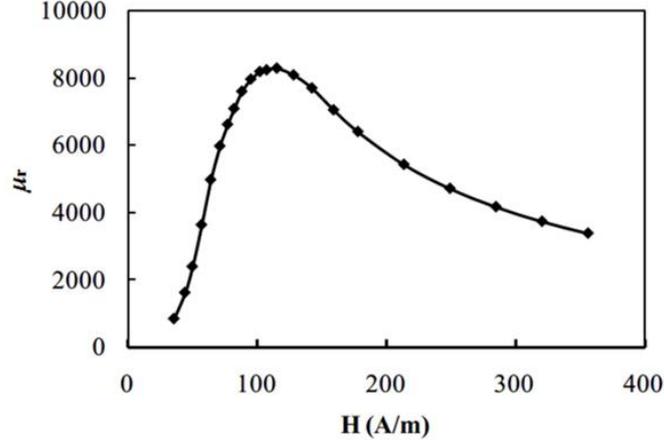

Fig. 4. Measured $\mu$-H curve of the ferromagnetic material (DT4C).

A preliminary test shows that coil heating is substantially reduced in the system described above. When the magnetic field in the air gap is 80 mT, the heating power of the exciting coil is 7.5 W. The temperature raise is 3 K for the exciting coil and 1 K in the air gap. In addition, for this structure the magnetic field in the gap mainly depends on the magnetic motive force and the shape of gap. The deformation of the exciting coil due to the heating has very small effect on the magnetic field in the gap. Thus the problem from the coil heating is improved greatly.

### III. ADOPTING THE MAGNETIC FLUX LINKAGE DIRECTLY

In joule balance, the mutual inductance between the exciting coils and the suspended coil is measured and becomes the key point [1]. The basic equation of the joule balance method is written as

$$[M(z_2) - M(z_1)]I_E I_S - mg(z_2 - z_1) = \int_{z_1}^{z_2} \Delta f(z) dz \qquad (1)$$

Where $M(z_1)$ and $M(z_2)$ are the mutual inductances between the mutual inductances between the suspended and exciting coils at positions $z_1$ and $z_2$ respectively; $m$ is the test mass; $g$ is the gravitational acceleration; $\Delta f(z)$ is the residual force at position $z$.

When the mutual inductance parameter is measured, the current in the exciting coils has to be changed and the non-linear characteristic and hysteresis as shown in Fig. 5 will bring a considerable measurement uncertainty. However, if we measure the difference of the magnetic flux linkage of the whole secondary coil, including the suspended coil and compensation coils, at two vertical positions directly to replace the mutual inductance parameter, this problem can be solved. If the magnetic flux linkage of the whole secondary coil is $\psi$, then equation (1) can be rewritten as

$$[\psi(z_2) - \psi(z_1)]I_S - mg(z_2 - z_1) = \int_{z_1}^{z_2} \Delta f(z) dz \qquad (2)$$

To calibrate the flux linkage $\psi(z_2) - \psi(z_1)$, let the suspended coil moves from $z_1$ to $z_2$. During the movement the current $I_E$ remains constant and the induced voltage of the suspended coil is

$$V(t) = \frac{d\psi(t)}{dt} \tag{3}$$

Integrating equation (3) from position $z_1$ to $z_2$, yielding

$$\psi(z_2)|_{t=t_2} - \psi(z_1)|_{t=t_1} = \int_{z=z_1, t=t_1}^{z=z_2, t=t_2} V(t)dt \tag{4}$$

Equation (4) shows that the quantity $\psi(z_2) - \psi(z_1)$ in (2) can be obtained directly by integrating the induced voltage in the whole secondary coil. The right hand of equation (4) is the area of function $V(t)$ during the movement of the suspended coil from position $z_1$ to $z_2$. It can be noticed that the area of the waveform of $V(t)$ is only dependent on the value difference of the flux linkage of the whole secondary coil at the position $z_1$ and $z_2$, i.e., $\psi(z_2) - \psi(z_1)$. The positions $z_1$ and $z_2$ are two static positions of the suspended coil. Especially, when the laser based position locking technique proposed in [6] is used, the position $z_1$ and $z_2$ can be very stable. Therefore, the value of $\psi(z_2) - \psi(z_1)$ is a fixed value when the current in the whole second coil remains constant. It is concluded from equation (4) that the value of the waveform area of $V(t)$ can be determined precisely and independent to the practical moving procedure of the suspended coil from $z_1$ to $z_2$. This is an advantage of the joule balance method. As an example, it is unnecessary to keep a constant moving velocity of suspended coil and the practical driving mechanism becomes simpler. At the same time, it is valuable to point out that the equation (2) and (4) can also be obtained easily from the basic equation of watt balance method.

To measure the area precisely, the standard square wave compensation method proposed for the mutual inductance measurement in [7] can be used. The waveform area of $V(t)$ is compared with a standard square wave, whose area can be known accurately by using a Programmable Josephson Voltage Standard (PJVS). At the same time, if an integrator is adopted to measure the area difference between $V(t)$ and the standard wave, the area of $\int_{t_1}^{t_2} V(t)dt$ can be determined precisely.

If we start from equation (2), the permanent magnet can also be used for the joule balance method. But in the design of Fig. 2, two coils with current are still used as the magnetic exciting source, because using coil will bring good flexibility in practical measurement. For example, the exciting current can be changed easily, thus an optimum working condition may be selected.

Although the coil heating is greatly reduced, the coil heating is still a disadvantage of this system. The 7.5 W coil heating power will lead to 3 K temperature raise in the exciting coil and 1 K in the gap in air condition. If the system is moved into vacuum, some special approaches for heat transmission should be considered.

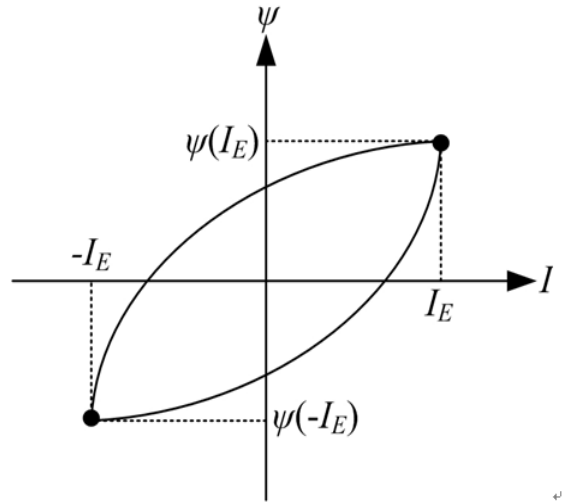

Fig. 5. The nonlinear characteristic and hysteresis in the mutual inductance measurement phase. $I_E$ denotes the current in the exciting coils and $\psi(I_E)$ is the flux linkage through the suspended coil.

## IV. MAGNETIC PROFILE

The designed parameters of the electromagnet are shown in Tab. 1.

TABLE I
PARAMETERS OF THE DESIGNED ELECTROMAGNET

| Quantity | Symbol | Value |
| --- | --- | --- |
| Inner yoke radius | $r_1$ | 160.0 mm |
| outer yoke radius | $r_2$ | 178.5 mm |
| Air gap length | $\delta$ | 18.5 mm |
| Air gap height | $h$ | 300 mm |
| Upper exciting coil turns and resistance | $N_1, R_1$ | 2608 turn, 18.5 Ω |
| Lower exciting coil turns and resistance | $N_2, R_2$ | 2608 turn, 18.5 Ω |
| Suspended coil turns and resistance | $N, R$ | 7136 turn, 4.1 kΩ |
| Upper compensation coil turns and resistance | $N_3, R_3$ | 3568 turn, 1.7 kΩ |
| Lower compensation coil turns and resistance | $N_4, R_4$ | 3568 turn, 1.7 kΩ |
| Exciting coil current | $I_E$ | 500 mA |
| Suspended coil current | $I_S$ | 7 mA |
| Compensation coil current | $I_C$ | 7 mA |

When the flux linkage is measured following equation (4), current is passing through the exciting coil only. The magnetic motive force of the primary coils, i.e., the exciting coils, $N_1 I_E - N_2 I_E$, is neutralized in the closed ferromagnetic path. The flux goes through the small loops shown in Fig. 2 (a), generating a horizontal magnetic flux density in the air gap.

For watt balances, the permanent magnet [8]-[15] or a superconducting solenoid [3] is employed, and hence the magnetic field generated is relative strong, e.g., 0.5 T~1 T. The profiles of these watt balances are conventionally changing several parts in $10^4$ in a typical measurement interval of several centimeters. In our design, since coils are applied as the excitation of the field, the magnitude of the magnetic field in the air gap is weaker than that of a watt balance, but a rather larger measurement interval can be obtained. Fig. 6 shows the calculated magnetic flux density in the air gap center, $B_r$, as a function of the vertical position $z$ ($z = 0$ mm is set at the vertical center of the air gap). It can be seen that the average value of $B_r$ in the range of [- 60 mm, 60 mm] is about 85.6 mT. Compared with the magnetic flux density of 6 mT generated from the compacted coil design in Fig. 1 (b), the magnetic field is increased by a factor ≈ 14. Besides, a flat magnetic profile with a height of more than 10 cm is obtained. The relative change of the magnetic profile in ± 60 mm is about 3 parts in $10^4$.

In the following, the effect of compensation coil in weighing phase will be discussed further. From Fig. 3 it seems that after compensation the distribution of magnetic field in Fig. 3 (c) is close to the original field in Fig. 3 (a). In fact, there is still small difference of the field distribution between Fig. 3 (a) and Fig. 3 (c). It should be considered for the precise measurement. The small change, expressed in forms of magnetic force to current ratio, can be written as

$$\ell = \frac{F}{I} = \ell_0 (1 + \alpha I + \beta I^2 + ...) \quad (5)$$

where $F$ is the magnetic force, $I$ the current in both the suspended and compensation coils, and $\ell$ the geometrical factor of the suspended coil and $\ell_0$ is its value in flux linkage measurement phase, i.e., $I = 0$. $\alpha$ and

$\beta$ are the linear and quadratic coefficients. The linear part can be removed from a current reversal [3, 8]. However, the quadratic parts cannot be eliminated and would generate a bias error during the measurement. In the following section, we will discuss this quadratic effect in the designed coil magnet.

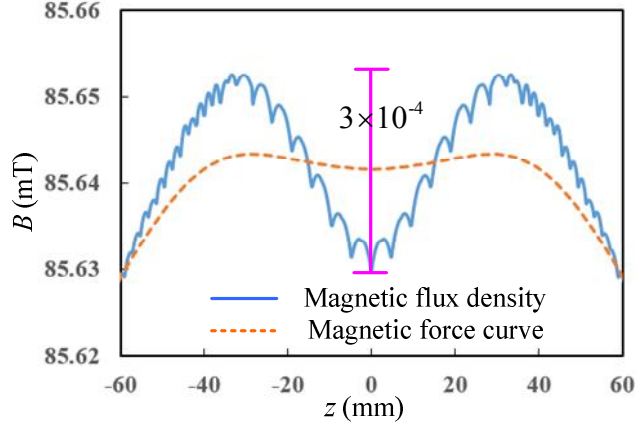

Fig. 6. The calculated magnetic flux density at air gap center as a function of vertical position. The relatively short spatial period of the magnetic flux density curve is caused by the low meshing fineness in the finite element method (FEM) calculation. The height of the suspended coil is 57 mm.

## V. Nonlinearity Error of the Ferromagnetic Material

There are mainly two aspects of the ferromagnetic material nonlinearity which can lead to the quadratic dependence of current $I$: the parallel component [16] and the perpendicular component [17]. In our design, the total magnetic motive force, $(N_1 - N_2) I_E + (N_3 - N_4) I_S$, in the closed ferromagnetic circuit is zero and hence no flux runs through the large loop. As a result, the magnitude of the parallel flux component discussed in [16] is removed.

As discussed in [17], the perpendicular component effect of a watt balance, causing a bias $\xi$ for the Planck constant measurement, can be written in terms of

$$\xi \approx \frac{\varepsilon}{B^2 \delta^n \mu^3} \frac{\partial \mu}{\partial H} \quad (2 < n < 3) \tag{6}$$

In (6), $\varepsilon$ is a constant; $B$ is the magnetic flux density in the air gap; $\delta$ is the air gap length; $\mu$ is the working permeability of the yoke; $\partial \mu / \partial H$ is the $\mu \sim H$ derivative at the working point. Thus the bias ratio between joule balance and a typical watt balance is written as

$$r = \frac{\xi_j}{\xi_w} = \left(\frac{B_w}{B_j}\right)^2 \left(\frac{\delta_w}{\delta_j}\right)^n \left(\frac{\mu_w}{\mu_j}\right)^3 \frac{(\partial \mu / \partial H)_j}{(\partial \mu / \partial H)_w} \tag{7}$$

In the coil magnet for joule balance, $B_a \approx 0.085$ T, compared to that of a watt balance constructed by permanent magnet (e.g., 0.4 T - 0.5 T), is about 5 times smaller, i.e., $B_w/B_j = 5$; $\delta = 18.5$ mm is a normal value to a watt balance (8 mm – 30 mm) and here we assume $\delta_w = \delta_j$. It can be seen from Fig. 3 that the $\mu$ - $H$ slope of DT4C is about 8000/100 = 80 m/A, which compare to the steel used in watt balance 1000/400 = 2.5 m/A, is increased by a factor of 32. It is reasonable to assume that the magnetic field difference between the actual working point and the ideal point (where the permeability is the maximum) for the yoke is the same in watt and joule balances. In this case, $(\partial \mu / \partial H)_j / (\partial \mu / \partial H)_w$ is evaluated to be 32 and $\mu_w/\mu_j = 1/8$. Taking all these values into (7), $r$ is calculated to be 1.6. It has been evaluated the perpendicular component effect to a watt balance $\xi_w$ is about several parts in $10^{10}$ [17]. According to (7), the amplitude of bias effect $\xi_j$ for joule balance is also of the same magnitude when $r = 1.6$.

## VI. Reluctance Force

In the weighing phase of watt or joule balance, there is a kind of additional force raised from the whole secondary coil reluctance change, named the reluctance [5]. In the presented electromagnet, a same current is passing through both the suspended coil and the compensation coils in the weighing phase, and the reluctance force is expressed as

$$f(z) = \frac{1}{2}I^2\frac{\partial L_M}{\partial z} + I^2\frac{\partial M_s}{\partial z} + \frac{1}{2}I^2\frac{\partial L_C}{\partial z} \tag{8}$$

where $I$ is the current in the whole secondary coils; $L_M$ is the inductance of the suspended coil, $L_C$ the inductance of the compensation coils, and $M_S$ the mutual inductance between the suspended coil and the compensation coils.

During the measurement, the position of the compensation coils is fixed, i.e., $L_C$ is a constant and $\partial L_C/\partial z = 0$, thus no additional force is produced by the right third term of equation (8). Since there is a closed magnetic circuit for the yoke, the first and the second terms in equation (8) are difficult to be separated. The reluctance force would pull the coil into the center of an iron structure, where has the minimum energy, and is independent to the current direction for the suspension coil. Since the proposed magnetic structure is symmetrical about $z$ axis, the reluctance force is an odd function of $z$. Here a finite element method (FEM) simulation is used to evaluate the amplitude of the reluctance force. In the simulation, there is only current in the whole secondary coils, but no current in the primary coil. The permeability of the yoke is set to 8000, which is close to the actual yoke working status. The calculation result, in which the magnetic force is a function of vertical position, is shown in Fig. 7.

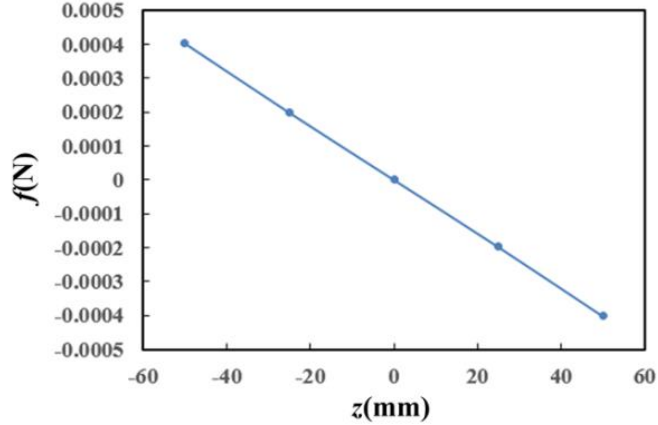

Fig. 7. The calculated reluctance force as a function of the vertical position.

It can be seen from Fig. 7 that the reluctance force $f(z)$ has a very good linearity. As $f(z)$ is an odd function, if a symmetrical vertical is chosen, i.e., $z_1 = -z_2$, the integral of $f(z)$ from $z_1$ to $z_2$ is zero. In reality, it is very difficult to make a small asymmetry for $z_1$ and $z_2$.

As shown in the simulation, the reluctance force error $\varsigma$ can be written as

$$\varsigma = \int_{z_1}^{z_2} f(z)dz = \int_{z_1}^{z_2} k(I_S^2)zdz = \frac{k(I_S^2)(z_2^2 - z_1^2)}{2} \tag{9}$$

where $k$ is the slope of the reluctance force curve shown in Fig. 7. Note that the reluctance is a force that is proportional to $I_S^2$ [18] and hence $k$ is proportional to $I_S^2$.

During the mass-on in the weighing mode, $I_{S1}$ is applied and we have

$$[\psi(z_2)-\psi(z_1)]I_{S1} - mg(z_2 - z_1) = \int_{z_1}^{z_2} \Delta f(z)dz + \frac{k(I_{S1}^2)(z_2^2 - z_1^2)}{2} \tag{10}$$

During the mass-off in the weighing mode, $I_{S2}$ is applied and we have

$$-[\psi(z_2)-\psi(z_1)]I_{S2} + mg(z_2 - z_1) = -\int_{z_1}^{z_2} \Delta f(z)dz + \frac{k(I_{S2}^2)(z_2^2 - z_1^2)}{2} \quad (11)$$

Subtracting (11) from (10), it yields the residual error $\eta$ that is expressed as

$$\begin{aligned}\eta &= \frac{[k(I_{S1}^2) - k(I_{S2}^2)](z_2^2 - z_1^2)}{4mg(z_2 - z_1)} \\ &\approx \frac{k(z_2 + z_1)}{4mg} \frac{(I_{S1}^2 - I_{S2}^2)}{I_S^2} \\ &\approx \frac{[f(z_2) - f(z_1)]}{4mg} \cdot \frac{z_2 + z_1}{z_2 - z_1} \cdot \frac{2(I_{S1} - I_{S2})}{I_S}\end{aligned} \quad (12)$$

In (12), the first term (force ratio) is calculated to be $4\times10^{-5}$; The second term (length ratio) can be 0.1 mm/100 mm = $1\times10^{-3}$; the third term (current ratio) is evaluated as $2\times10^{-3}$. Therefore, in theory, the error due to the reluctance can be controlled well below $1\times10^{-8}$. Note that the model presented in (12) is based on an ideal construction that the reluctance force is proportional to $z$. In practice, the $f(z)$ function should be measured and the residual error $\eta$ should be reevaluated according to experiment data.

## VII. Conclusion

The self-heating of coils used is a big uncertainty source for the joule balance method. A new coil set with ferromagnetic material is proposed to solve this problem. The effect from the nonlinear characteristic and hysteresis of the ferromagnetic material can be removed by measuring the difference of the magnetic flux linkage of the whole secondary coil directly to replace the mutual inductance parameter. The theoretical analysis shows that the nonlinear error is negligible compared with a typical uncertainty of several parts in $10^8$, and the reluctance force can be well removed by means of mass on and mass off.